\documentclass[a4paper,11pt]{article}
\usepackage[margin=1.4cm]{geometry}
\usepackage[utf8]{inputenc}
\usepackage{graphicx}
\usepackage{xcolor}
\definecolor{darkgreen}{rgb}{0,0.5,0}
\usepackage[bookmarks,bookmarksopen,pdfpagemode=UseOutlines,pdfnewwindow=true,
  hyperfootnotes=false,colorlinks,urlcolor=blue,citecolor=darkgreen]{hyperref}
\usepackage{amsmath,amssymb,amsfonts,tabularx,multicol,multirow,rotating,nicefrac}
\usepackage{dcolumn} 
\newcolumntype{d}[1]{D{.}{.}{#1}} 

\makeatletter
\newcommand{\mythanks}[1]{\hbox{\@textsuperscript{\normalfont#1}}}
\makeatother

\author{
     Pablo Jensen\mythanks{1,2,3,4,*}
\and Jean-Baptiste Rouquier\mythanks{1,2,6}
\and Pablo Kreimer\mythanks{5}
\and Yves Croissant\mythanks{1,4}}
\begin{document}
\footnotetext[1]{Universit\'e de Lyon}
\footnotetext[2]{Institut des Syst\`emes Complexes Rh\^one-Alpes (IXXI)}
\footnotetext[3]{Laboratoire de Physique, \'Ecole Normale Sup\'erieure de Lyon and CNRS, 69007 Lyon, FRANCE}
\footnotetext[4]{Laboratoire d'\'Economie des Transports, Universit\'e Lyon 2 and CNRS, 69007 Lyon, FRANCE}
\footnotetext[5]{Instituto de Estudios de la Ciencia, Solis 1067, C-1078 AAU Ciudad de Buenos Aires, Argentina}
\footnotetext[6]{Laboratoire d'Informatique du parall\`elisme, \'Ecole Normale Sup\'erieure de Lyon and CNRS, 69007 Lyon, FRANCE}
\renewcommand{\thefootnote}{\fnsymbol{footnote}}
\footnotetext[1]{Corresponding author : pablo.jensen@ens-lyon.fr}

\title{\textbf{Scientists who engage with society perform better academically}}
\date{July 1, 2008}
\maketitle

\begin{abstract}
Today, most scientific institutions acknowledge the importance of opening the so-called ``ivory tower'' of academic research through popularization, industrial innovation or teaching. However, little is known about the actual openness of scientific institutions and how their proclaimed priorities translate into concrete measures. This paper helps getting an idea on the actual practices by studying three key points: the proportion of researchers who are active in dissemination, the academic productivity of these active scientists, and the institutional recognition of their activity in terms of careers. This paper answers these questions by analyzing extensive data about the academic production, career recognition and teaching or public/industrial outreach of several thousand CNRS scientists from many disciplines. We find that, contrary to what is often suggested, scientists active in dissemination are also more active academically. However, their dissemination activities have almost no impact (positive or negative) on their career.
\end{abstract}

\section{Introduction}

Researchers and academic institutions seem to have admitted the importance of establishing strong ties between science and society. In the United Kingdom, Martin Rees, president of the Royal Society, points out that ``Researchers need to engage more fully with the public. The Royal Society recognizes this, and is keen to ensure that such engagement is helpful and effective''. A recent survey carried out by the Royal Society finds that ``Most researchers have highlighted that social and ethical implications exist in their research, agree that the public needs to know about them, and believe that researchers themselves have a duty, as well as a primary responsibility, for communicating their research and its implications to the non-specialist public.''~\cite{royal2006}

In France, the CNRS declares in the document supposed to steer his long-term policy, the ``Multi-year action plan''~\cite{CAPCNRS}, that one of the six top priorities is ``to transfer research results to industries'' and another ``to strengthen the relations between science and society''. In February 2007, CNRS organized an official workshop on ``Science and Society in transformation'', in presence of many CNRS officials~\cite{SSCNRS}. This attitude seems to be shared by the majority of researchers: in her study on the attitudes of researchers towards popularization~\cite{cheveigne}, Suzanne de Cheveign\'e concluded: ``All interviewed researchers unanimously declared: popularization is now a key and unavoidable component of research work.'' Motivations provided by researchers are numerous: the yearning to inform the public, to make one's field of research better known and encourage students to take up science, or the need to account to civil society for the use of funds provided to laboratories.

The reality on the field is generally aloof from these generous ideas. For example, in the CNRS report for candidating to the ``Directeur de Recherche'' (Senior Scientist) position, a mere 9 lines are provided to summarize twenty years of research dissemination. Likewise, the Royal Society survey concludes that, for most scientists, ``research is the only game in town'', and popularization has to be done after one is through with ``real'' work. 

The purpose of this paper is to obtain an empirical picture of dissemination practices in CNRS.
We have presented in a previous study~\cite{jcom} a statistical view of scientists involved in popularization.  Here, we also include data on teaching and industrial collaborations. Moreover, we correlate these data with scientists' academic activity, as quantified by bibliometric records. Therefore, we are able, for the first time, to answer two important questions about scientists active in dissemination: are they ``bad scientists'' as some scientists suggest~\cite{royal2006}? Do they get any institutional recognition in terms of careers ? We answer these questions by analyzing extensive data about the academic production, career recognition and teaching or public/industrial outreach of several thousand CNRS scientists from many disciplines.

\section{Methodology}

Thanks to the CNRS Human Resources Direction, we have gathered data on the dissemination activities (public outreach, industrial collaborations and teaching) of the 11\,000 CNRS scientists over a three-year period (2004-2006). It should be noted that these data are {\it declared} by scientists in their annual report (``Compte Rendus Annuels des Chercheurs'' or CRACs). This annual report is not judged very important for the career, serious evaluations taking place only when scientists candidate to senior positions. However, filling out the report is mandatory and most researchers (over 90\% each year) do fill it in due time. Many reasons could lead to some underestimation of the amount of activities declared, including fear of misperception of popularization activities by commitees, laziness to report faithfully these minor activities, etc. Inversely, lack of control of these items could favor some overreporting of dissemination activities, although this is not likely since they have almost no perceived impact on career. Hence, we could anticipate some underestimation of dissemination activities in the figures below.

``Popularization'' activities include public or school conferences,
interviews in newspapers, collaboration with associations... Clearly,
there is no entirely satisfactory definition of popularization. As
Stephen Hilgartner convincingly shows~\cite{hilgartner}, there exists
in fact a continuous gradation going from technical litterature to
popular science, with no clear cut indicating where popularization
begins. Here, popularization actions are declared by scientists
themselves, according to the following operational criterion :
popularization means wide audience, actions aiming at non specialized
public. For more details, we refer the reader
to Appendix 1 and reference~\cite{jcom}. ``Industrial collaborations'' mainly means contracts
with industrial partners or funding from non academic sources
(regional, specific programs \ldots). For natural sciences, contracts
with industrial partners dominate, while funding from specific
programs accounts for most of such actions in the social and human
sciences.  Both these types of actions account for 81\% of the
actions, while patents represent 16.5\% and licenses 2.5\% of the
actions. ``Teaching'' is only characterized by the annual number of hours
dedicated to this activity. CNRS researchers have no teaching duties.

We have described previously~\cite{testh} how we manage to obtain a large but robust database of bibliometric indicators for the CNRS scientists. Briefly stated, our method uses the ``Author search'' of~\cite{wos} on the subset of 8750 scientists having filled out the CNRS report the last three years. We exclude researchers in Social Sciences (their bibliographic record is not well documented in WoS) and in High Energy Physics (too few records in the CNRS database), leading to 6900 names. After filtering records suspected to be erroneous, we obtain a database of 3659 scientists with reliable bibliometric indicators, as checked by close inspection of several hundred records and good prediction of scientists promotions. A more detailed description of our method is given in Appendix 2 and reference~\cite{testh}.

We have used several bibliometric indicators as proxies for academic
activity: number of papers published, number of papers published per
year, number of citations or Hirsch index~\cite{hirsch}. It could be
argued that $h$ is not a good measure when comparing the scientific
activity of researchers with very different career lengths, because it
automatically increases with time. A more relevant measure might be
$h$ divided by the career length in years $h_y$~\cite{hirsch},
although we have shown that it is not perfect either, since its
average value for CNRS scientists decreases with scientist's
age~\cite{testh}. However, since $h_y$ is closer to a constant than
$h$ for scientists with different career lengths~[ibid.], we will
use it, along with the average number of papers published per year
(which is indeed constant over scientists' ages) and other
bibliometric indicators.

\section{Proportion of active scientists}

\begin{figure*}[hp]
\begin{center}
\includegraphics[width=0.5\textwidth]{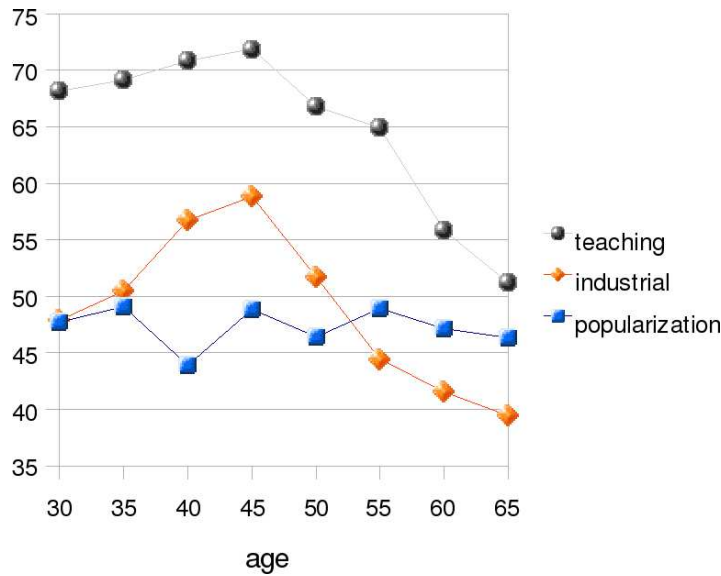}
\caption{Evolution of the proportion of scientists active in dissemination as a function of their age. Data correspond to the whole database, i.e. before filtering with bibliometric indicators.} 
\label{fig:agedissem}
\end{center}
\end{figure*}

\begin{figure*}[hp]
\begin{center}
\includegraphics[width=0.5\textwidth]{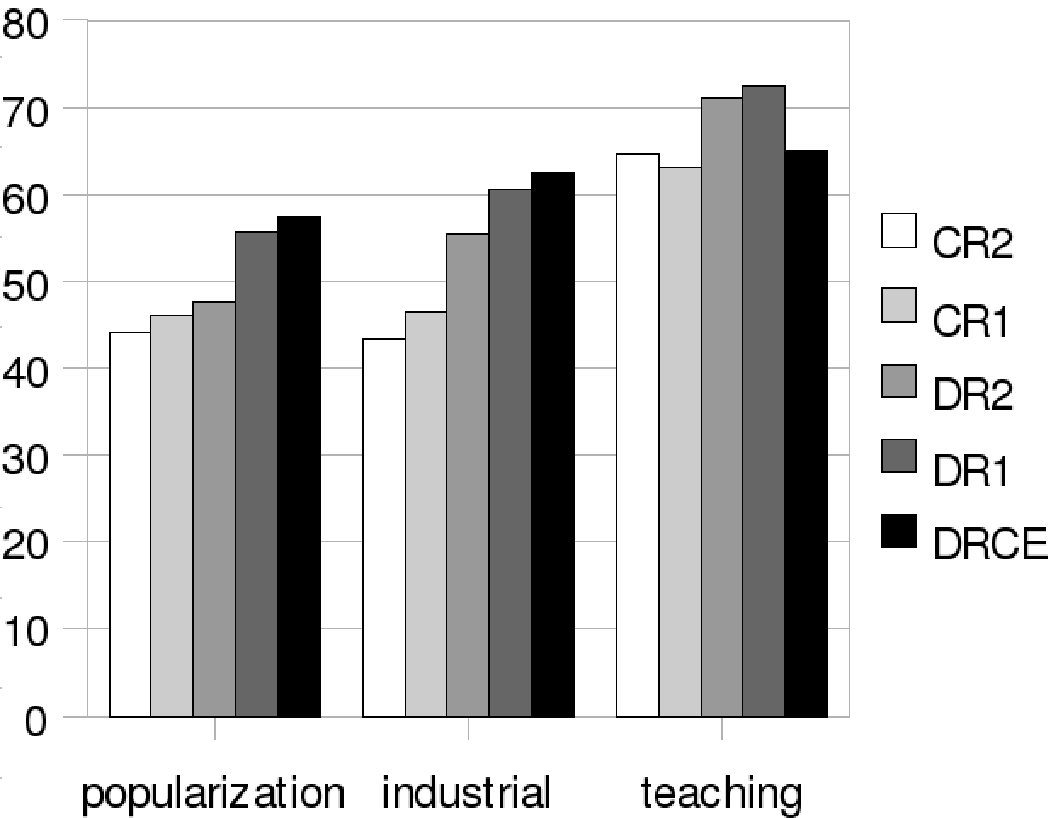}
\caption{Evolution of the proportion of scientists active in dissemination as a function of their position. Data correspond to the whole database, i.e. before filtering with bibliometric indicators.}
\label{fig:gradedissem}
\end{center}
\end{figure*}

A summary of the subdisciplines encompassed by our database, together with some characteristic average values, is shown in tables~\ref{propDS} and~\ref{propsec}. In Figures~\ref{fig:agedissem} and~\ref{fig:gradedissem}, we show the proportion of active scientists by age and grade, for each of the three dissemination activities. Overall, CNRS scientists carry out more than 7\,000 popularization actions and more than 4\,000 industrial collaborations per year, and these figures are increasing. Table~\ref{propDS} shows however that the activity is very unequally distributed: over a three-year period, about half of scientists remained inactive in popularization or industrial collaborations. For more details on popularization activities, see our previous analysis~\cite{vulganature,jcom}.

These large scientific domains are in fact heterogeneous. It is interesting to study more disaggregated data, at the discipline level (corresponding to CNRS scientific ``sections''). Table~\ref{propsec} shows in detail the proportion of active scientists in each of the subdisciplines for popularization, teaching and industrial collaborations.

\begin{table}[p]
\begin{center}
\caption{Percentage of inactive (no action or no teaching respectively), active (less than 10 outreach actions or less than 4 industrial collaborations or less than 210 teaching hours  respectively) and very active scientists for the CNRS scientific domains. This division in subpopulations is more instructive than the mean number of actions, as the activity is very unequally distributed among researchers: the 5\% most active account for half of the actions~\cite{vulganature}. Figures correspond to the activity cumulated over 2004-2006.  CNRS researchers have no teaching duties.} 
\begin{tabular}{|l@{ }||
c|c|c||c|c|c||c|c|c|}
\hline
 & \multicolumn{3}{@{ }c@{ }||}{Popularization}
 & \multicolumn{3}{@{}c@{}||}{\begin{tabular}{c}Industrial\\Collaboration\end{tabular}}%
 & \multicolumn{3}{c@{ }|}{Teaching} \\
\multicolumn{1}{|p{2cm}||}{DOMAIN}
  & \rotatebox{90}{inactive} & \rotatebox{90}{active} & \rotatebox{90}{\begin{tabular}{@{}l}very\\ active\end{tabular}}
  & \rotatebox{90}{inactive} & \rotatebox{90}{active} & \rotatebox{90}{\begin{tabular}{@{}l}very\\ active\end{tabular}}
  & \rotatebox{90}{inactive} & \rotatebox{90}{active} & \rotatebox{90}{\begin{tabular}{@{}l}very\\ active\end{tabular}}\\
\hline\hline
Physical sciences              & 59  &  39  & \hphantom12 & 63 & 35 & 2 & 46 & 50 & 4 \\ 
High Energy Physics            & 45  &  54  & \hphantom11 & 95 & 5  & 0 & 71 & 27 & 2 \\
Life sciences                  & 64  &  34  & \hphantom12 & 43 & 53 & 4 & 33 & 66 & 1 \\
Engineering                    & 52  &  46  & \hphantom12 & 19 & 74 & 7 & 22 & 69 & 8 \\
Chemistry                      & 65  &  34  & \hphantom11 & 39 & 52 & 9 & 42 & 55 & 2 \\
Earth Sciences, Astrophysics   & 36  &  57  & \hphantom17 & 59 & 41 & 1 & 31 & 67 & 1 \\
Social Sciences                & 27  &  62  &          10 & 68 & 32 & 0 & 17 & 76 & 8 \\
\hline
All CNRS                       & 53  &  43  & \hphantom14 & 49 & 47 & 4 & 33 & 63 & 4 \\
\hline
\end{tabular}
\label{propDS}
\end{center}
\end{table}

\begin{table}[p]
\begin{center}
\small
\caption{Details of Table~\ref{propDS} by subdiscipline. The precise names of the CNRS ``sections'' have been shortened for simplicity. The discipline ``High Energy Physics'' of Table~\ref{propDS} corresponds to ``Interactions, particles \& strings''. As in Table~\ref{propDS}, we show the percentage of active (less than 10 popularization actions or less than 4 industrial collaborations or less than 210 teaching hours  respectively) and very active scientists (i.e. more active than the previous figures) for the CNRS scientific subdisciplines. For simplicity, we have not shown the percentages of inactive (no action or no teaching respectively) scientists. These can be easily calculated from the difference to 100\% of the sum of the two columns ``active'' and ``very active''. Figures correspond to the activity cumulated over 2004-2006.} 
  \centerline{\begin{tabular}{|crp{4.89cm}|cd{2}|cd{2}|cd{3}|}
    \hline
&&&\multicolumn{2}{c|}{popularization}&\multicolumn{2}{c|}{industrial}&\multicolumn{2}{c|}{teaching}\\
    &&Subdiscipline
    &active
    &\multicolumn{1}{c|}{\parbox{0.9cm}{\rule{0pt}{1eM}very active\rule[-1ex]{0pt}{1pt}}}
    &active
    &\multicolumn{1}{c|}{\parbox{0.9cm}{very active}}
    &active
    &\multicolumn{1}{c|}{\parbox{1.18cm}{very active}}
    \\
    \hline
    \multirow{6}{*}{\parbox{1cm}{\centering Physical\newline sciences}}
    &1&Mathematics&30&3&16&1&61& 7 \\
    &2&Physics, theory \& method&29&3&25&1&43& 3 \\
    &3&Interactions, particles \& strings&54&1&5&0&27& 2 \\
    &4&Atoms \& molecules&43&1&40&3&51& 4 \\
    &5&Condensed matter: dynamics&44&3&45&2&51& 3 \\
    &6&Condensed matter: structure&42&1&42&1&38& 6 \\
    \hline
    \multirow{4}{*}{Engineering}
    &7&Information science&47&1&71&3&76& 6 \\
    &8&Micro \& nano-technologies&44&2&72&15&59& 11 \\
    &9&Materials \& structure&48&2&80&3&76& 14 \\
    &10&Fluids \& reactants&45&4&80&5&68& 6 \\
    \hline
    \multirow{6}{*}{Chemistry}
    &11&Super/macromolecular systems&42&1&52&11&62& 1 \\
    &12&Molecular architecture synthesis&29&0&47&8&54& 4 \\
    &13&Physical chemistry&33&1&43&3&53& 2 \\
    &14&Coordination chemistry&37&1&53&14&44& 3 \\
    &15&Materials chemistry&35&1&59&12&57& 1 \\
    &16&Biochemistry&28&1&52&7&66& 0 \\
    \hline
    \multirow{4}{*}{\parbox{3cm}{\centering Earth sciences,\newline astrophysics}}
    &17&Solar systems \& the universe&56&14&20&0&44& 1 \\
    &18&Earth \& earth plants&59&4&39&0&79& 3 \\
    &19&Earth systems: superficial layers&56&5&52&1&64& 1 \\
    &20&Continental surface&56&3&66&1&80& 1 \\
    \hline
    \multirow{10}{*}{Life sciences}
    &21&Molecular basis of life systems&27&1&58&6&61& 3 \\
    &22&Genomic organization&26&1&43&2&62& 2 \\
    &23&Cellular biology&25&0&50&4&63& 2 \\
    &24&Cellular interaction&29&0&50&5&66& 2 \\
    &25&Physiology&32&2&57&3&64& 0 \\
    &26&Development, evolution&32&1&46&2&67& 0 \\
    &27&Behavior, cognition \& brain&59&6&63&1&78& 1 \\
    &28&Integrative vegetal biology&34&0&52&4&60& 0 \\
    &29&Biodiversity, evolution&54&8&60&1&81& 1 \\
    &30&Therapy, pharmacology&38&1&58&12&63& 1 \\
    \hline
    \multirow{10}{*}{\parbox{3cm}{\centering Human \& social\\ sciences}}
    &31&Human evolution&63&16&24&0&82& 5 \\
    &32&Ancient \& medieval history&66&9&13&0&71& 8 \\
    &33&Modern \& contemporary history&65&10&19&0&75& 5 \\
    &34&Languages, language \& speech&54&4&32&1&75& 8 \\
    &35&Philosophy&58&10&20&1&63& 8 \\
    &36&Sociology&68&11&42&0&80& 9 \\
    &37&Economics \& management&43&6&61&1&80& 7 \\
    &38&Society \& cultures&69&9&26&0&78& 2 \\
    &39&Environment \& society&71&9&62&0&82& 11 \\
    &40&Politics, power&64&19&49&1&79& 15 \\
\hline
\end{tabular}}
\label{propsec}
\end{center}
\end{table}

\begin{table}[ht]
\begin{center}
\caption{Percentage of active scientists in popularization, industrial collaborations or teaching as a function of their age (activity cumulated over 2004-2006, whole database including Social Sciences and High Energy Physics). CNRS researchers have no teaching duties.}
\begin{tabular}{rrrr}
\hline
age & pop & indus & teach \\
\hline
$<$30 & 47.7 & 47.9 & 68.1 \\
35 & 49.1 & 50.4 & 69.1 \\
40 & 43.9 & 56.7 & 70.9 \\
45 & 48.8 & 58.8 & 71.9 \\
50 & 46.4 & 51.7 & 66.8 \\
55 & 48.9 & 44.4 & 64.9 \\
60 & 47.1 & 41.5 & 55.8 \\
$>$65 & 46.3 & 39.4 & 51.2 \\
\hline
\end{tabular}
\label{tab:agedissem}
\end{center}
\end{table}

\begin{table}[ht]
\begin{center}
\caption{Percentage of active scientists in popularization, industrial collaborations or teaching as a function of their position (activity cumulated over 2004-2006, whole database including Social Sciences and High Energy Physics).  The different positions of CNRS scientists are, by increasing hierarchical importance: ``Charg\'e de Recherche 2$\null^\text{e}$ classe'' (CR2), ``Charg\'e de Recherche 1$\null^\text{re}$ classe'' (CR1), ``Directeur de Recherche 2$\null^\text{e}$ classe'' (DR2), ``Directeur de Recherche 1$\null^\text{re}$ classe'' (DR1) and ``Directeur de Recherche de Classe Exceptionnelle'' (DRCE). The proportion of CNRS scientists for each position are given in the last column. CNRS researchers have no teaching duties.}
\begin{tabular}{rrrrr}
\hline
position & pop & indus & teach & \%\\
\hline
CR2 & 44.1 & 43.3 & 64.7 & 5.9\\
CR1 & 46.1 & 46.4 & 63.2 & 52.9\\
DR2 & 47.5 & 55.5 & 71.0 & 31.5\\
DR1 & 55.7 & 60.6 & 72.4 & 8.8\\
DRCE & 57.5 & 62.5 & 65.0 & .9\\
\hline
\end{tabular}
\end{center}
\label{tab:gradedissem}
\end{table}

\section{Academic achievement of open scientists}

A large fraction of scientists view dissemination activities as a low status occupation, done by ``those who are not good enough for an academic career''~\cite{royal2006}. This common perception is captured by the well-known ``Sagan effect'': popularity and celebrity with the general public are thought to be inversely proportional to the quantity and quality of real science being done~\cite{sagan}. Sagan's biographers~\cite{shermer} have shown that Harvard's refusal of Sagan's bid for tenure, and the National Academy of Science's rejection of the nomination of Sagan for membership, was a direct result of this perception. By analyzing his publication record, they have also shown that there is no such effect: ``Throughout his career, which began in 1957 and ended in December 1996, upon his untimely death, Sagan averaged a scientific peer-reviewed paper per month. The `Sagan Effect', at least when applied to Sagan himself, is a Chimera''~\cite{shermer}. In the following, we will test on a larger scale whether such an effect exists for the CNRS scientists, i.e. whether scientists engaged in dissemination are academically less active than average. To anticipate our conclusion, let us say that we find exactly the opposite correlation: scientists connected with society are more active than average, for reasons we then investigate.

\subsection{Comparing bibliometric indicators of active and inactive scientists}

We begin by comparing the average academic activity of active (in dissemination) and inactive scientists. The precise question we investigate is: if we choose randomly a scientist and ask her whether she is active in dissemination, does the answer tell us something about her academic activity? According to the common view quoted above, the answer should be that an active scientist has, on average, a weaker academic activity, which should correspond to lower bibliometric indicators. Our data shows exactly the opposite effect.

\subsubsection{Comparing bibliometric indicators of active and inactive scientists}

\begin{figure*}[hp]
  \begin{center}
\includegraphics[width=0.5\textwidth]{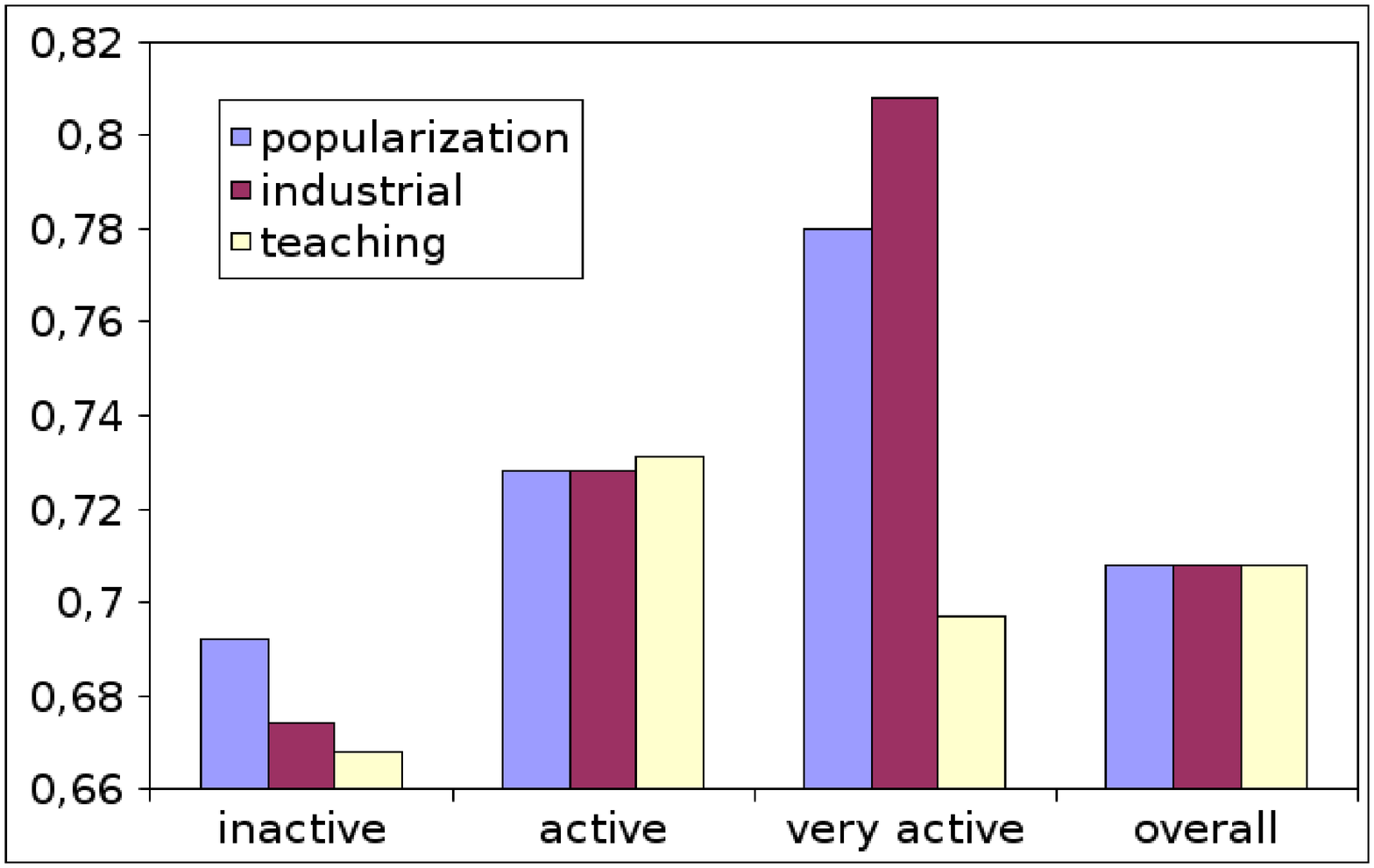}
\caption{Average $h_y$ for inactive, active, or very active scientists (see Table~\ref{propDS} for the definitions) in the different dissemination activities. Again, we exclude researchers in Social Sciences because their bibliographic record is not well documented in WoS. Variance tests
on the indicators ensures that they are strongly significant (for popularization: $F = 6.9$, p-value = 0.01; for industrial collaborations: $F = 18.6$, p-value = 0.00004. For teaching, active scientists have a significantly higher $h_y$ than the non active, p-value = 0.0003). However, contrary to dissemination, the very active ones have the same $h_y$ than the mean (the small difference is not statistically significative). Our data point to an ``optimal'' value of roughly 20 - 30 teaching hours per year, additional hours lowering $h_y$.}
\label{hy_activens}
\end{center}
\end{figure*}

Figure~\ref{hy_activens} shows that activity in dissemination is
correlated with higher academic indicators. Scientists inactive in
both popularization and industrial collaborations (roughly 30\%) have
a lower academic activity ($h_y=0.65$), which still decreases for the
ones also inactive in teaching (15\%, $h_y=0.62$). If one uses the
number of papers published per year, the conclusion is similar: the
average value is 2.28, while dissemination active scientists have
significantly higher average values (popularization: 2.38, p-value
$2.6 10^{-5}$, industrial collaboration 2.45, p-value $<2.2 10^{-16}$,
teaching: 2.35, p-value $8 10^{-6}$).

A potential danger of this kind of general comparison is the well-known variability of average $h$ indexes among different scientific disciplines~\cite{iglesias}. Therefore, we calculated the differences in $h_y$ between scientists of the same discipline but different dissemination activities. Our results (Table~\ref{hyDS}) confirm the fact that open scientists are always, on average, academically more active that the inactive ones, even if the smaller number of scientists investigated prevents the results from being statistically significative for some disciplines. Note also that $h_y$ is not always the best indicator of academic activity: for Engineering, the average number of papers published accounts better for promotion~\cite{testh}. Taking this indicator, differences between active and inactive scientists become significative in favour of active scientists (for example, popularization active scientists from Engineering sciences have an average publication rate of 2.12, instead of 1.93 for inactive ones, p-value = 0.07).

\begin{table}[hp]
\begin{center}
\caption{Differences in scientific activity --- as measured by the normalized Hirsch index --- for different subpopulations, characterized by the strength of their dissemination activities. To simplify the presentation, we only keep two categories: inactive and active, the latter grouping the ``active'' and ``very active'' categories of Table~\ref{propDS}. We also show the p-values obtained by a standard ``Welch Two Sample t-test'' and the number of scientists in each domain.}
\begin{tabular}{|l|d{5}|d{4}|d{4}@{}l|c|}
\multicolumn{6}{c}{Popularization}\\
\hline
 & \multicolumn{1}{c|}{inactive $h_y$} & \multicolumn{1}{c|}{active $h_y$}
 & \multicolumn{2}{c|}{p-value} & \multicolumn{1}{c|}{Number of scientists} \\
\hline
Physical sciences
        &        0.68        & 0.73        &        0.036& * & 669        \\
Life sciences 
        &        0.75        & 0.81        & 0.0018& **         & \llap1275 \\        
Engineering 
        &        0.50                &        0.52        & 0.38& & 504\\
Chemistry 
        & 0.73        &         0.74        &        0.54 && 848\\
Earth Sciences, Astrophysics 
&         0.69        & 0.77        &  0.037& * & 363                \\
\hline
\multicolumn{6}{c}{}\\
\multicolumn{6}{c}{Industrial Collaboration}\\
\hline
 & \multicolumn{1}{c|}{inactive $h_y$} & \multicolumn{1}{c|}{active $h_y$} & \multicolumn{2}{c|}{p-value} & Number of scientists \\
\hline
Physical sciences
        &        0.65        & 0.78        & < 1.E-6        &*** & 669        \\
Life sciences 
        &        0.69        & 0.83        & < 1.E-6        &*** & \llap1275 \\        
Engineering 
        &        0.47                &        0.52        & 0.17 && 504\\
Chemistry 
        & 0.69        &         0.75        &        0.0066 &** & 848\\
Earth Sciences, Astrophysics 
&         0.74        & 0.74        &  0.96 & &363                \\
\hline
\multicolumn{6}{c}{}\\
\multicolumn{6}{c}{Teaching}\\
\hline
 & \multicolumn{1}{c|}{inactive $h_y$} & \multicolumn{1}{c|}{active $h_y$} & \multicolumn{2}{c|}{p-value} & Number of scientists \\
\hline
Physical sciences
        &        0.69        & 0.70        &        0.77 && 669        \\
Life sciences 
        &        0.67        & 0.81        & < 1.E-6        &*** & \llap1275 \\        
Engineering 
        &        0.47                &        0.52        &  0.12 && 504\\
Chemistry 
        & 0.69        &         0.76        &         0.0004
     &*** & 848\\
Earth Sciences, Astrophysics 
&         0.74        & 0.76        &  0.58 && 363                \\
\hline
\end{tabular}
\label{hyDS}
\end{center}
\end{table}

\subsubsection{Scientists active in all dissemination activities}

It is also interesting to look at the academic records of the scientists active in all three dissemination actions. They represent roughly 20\% of our 3659 database, which is much more than expected if the engagements in the three different dissemination activities (teaching, industrial collaboration and popularization) were uncorrelated (14 \%). This points to an ``open'' attitude,  which makes a scientist practicing popularization more prone to teach or establish industrial collaborations. This high percentage is contrary to what one could expect from a ``time consumption'' argument, where each of these activities lowers the activity in the others. From an academic point of view, scientists who cumulate more than one dissemination activity are more active academically than those who carry out only one of them. The precise values are: scientists active in all three dissemination activities have a $h_y$ of 0.75 against 0.70 for the others (p-value = 0.0001), those active in industrial collaborations and teaching 0.74 against 0.69 (p-value = 2.0 $10^{-5}$), those active in industrial collaborations and popularization 0.74 against 0.70 (p-value = 0.012), those active in popularization and teaching 0.74 against 0.70 (p-value = 0.0008).

\subsubsection{Dissemination activity of the ``best'' scientists}

One can also look at the dissemination attitude of the (academically) most active scientists, taken as those whose $h$ increases faster than their career time ($h_y > 1$, totalizing 1/6 of CNRS researchers). They are more active in both outreach (44\% of active instead of 37\%, p-value = 0.0035), industrial dissemination (56\% of active instead of 51\%, p-value = 0.035) and teaching (69\% of active instead of 60\%, p-value = 7.5 $10^{-5}$). The same correlations are found by discipline, even if, again, the differences are less significative, except for biology and physics.

\subsubsection{Dissemination activity of ``those who are not good enough''}

Finally, one can investigate whether ``those who are not good enough for an academic career''~\cite{royal2006} are the most active in dissemination. Our previous result suggest that this is not so, which is confirmed by a statistical analysis. Taking as ``not good enough'' the 25\% of CNRS scientists with the lowest $h_y$ (lower than .5), we find that these scientists are less active in dissemination, the precise figures being 39.4\% active for popularization instead of 41.4\% for the rest of the scientists (p-value = .25 i.e. non significative), 52.1\% active for industrial collaboration instead of 57.8\% (p-value = .0025) and 60\% active for teaching instead of 67\% (p-value = .0002). Even stronger differences (all highly significative statistically) are found if the number of publications per career year is used as the bibiliometric indicator, which could be more appropriate when comparing scientists of different ages~\cite{testh} (for example, 35.7\% of the ``not good enough'' are active for popularization instead of 42.6\% in the rest, p-value = .00021).

\section{Which scientists are active in dissemination?}

In the previous section, we have shown that scientists engaged in dissemination, be it popularization, teaching or industrial collaborations, are academically more active than inactive researchers. To be precise, we have shown that if you select randomly a scientist and ask him whether he is active in dissemination, a positive answer implies higher bibliometric indicators. We now investigate the separate effects of all the available scientists' characteristics on the probability that they are active in dissemination. This will help us to interpret the correlations between academic and dissemination activities (next section). 

We have conducted a statistical analysis intending to single out the
individual effects of each one of these characteristics, all other
things being equal. For example, for a (hypothetical) average
researcher, we analyze the effect of her position, i.e. how much it
separately increases (or diminishes) her probability of being active
in popularization. Since the variable that we investigate is a logical
variable (either active or inactive), we have used a standard logit
regression model~\footnote{The statistical analysis was carried out
with the open software ``R'' (http://www.r-project.org/)}. In this
model, the probability of being active is written as
$P(y_i=1)=F(\beta'x_i)$ where $\beta'$ is the vector of fitted
coefficients and $x_i$ the vector of characteristics of scientist $i$
(age, position...). The marginal effect of a variation of the variable
$x_{ik}$ (where $k$ refers to one of the characteristics) on the
probability of being active can be written as $\frac{\partial
P(y_i=1)}{\partial x_{ik}}=\frac{\partial F}{\partial
x_{ik}}(\beta'x_i)=\beta_k F'(\beta' x_i)$. In a logit model,
$F(z)=\frac{e^z}{1+e^z}$, which leads to
$F'(z)=\frac{e^z}{(1+e^z)^2}=F(z) (1-F(z))$. This function reaches its
maximum for $z=0$, which corresponds to an activity probabililty of
0.5, leading to a proportionnality coefficient of 1/4. Therefore, a
simple interpretation of the effect of a scientist characteristic on
its probability of being active is the following : the maximum
marginal effect of a characteristic equals the corresponding
coefficient divided by a factor 4. For example, the isolated effect of
an age increase of one year is a decrease of about $0.027/4*100 \simeq
.6\%$ of the probability of popularizing. Being ``CR1'' decreases the
probability of being active in industrial collaborations by 18\%
compared to a ``DR2'' sharing the same characteristics (age, sex,
subdiscipline...).

Our results are summarized in Tables~\ref{regression1}
and~\ref{regression2}. The main influences of scientists'
characteristics, all other things being equal, are:
\begin{enumerate}
\item {\it position}: as scientists reach higher positions, they become significantly more active in all dissemination activities
\item {\it academic record}: there is no significant influence except for industrial collaborations. For this activity, scientists with higher Hirsch index are more active
\item {\it age}: dissemination activities decrease with age
\item {\it gender}: women are more active in popularization, men in teaching, and there is no significative difference in industrial collaborations
\end{enumerate}

\begin{figure*}[hp]
\begin{center}
\includegraphics[width=0.5\textwidth]{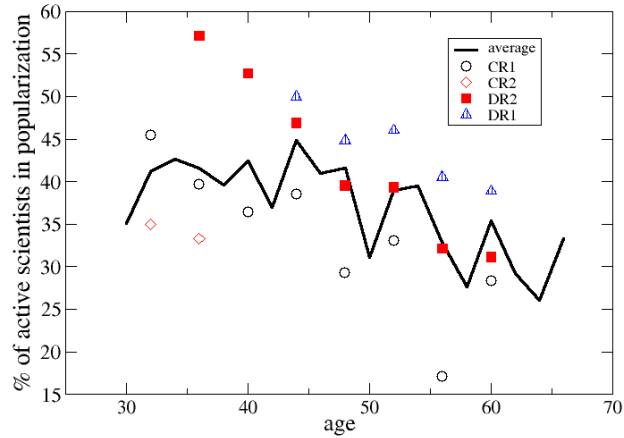}
\caption{Evolution of the proportion of scientists active in popularization as a function of their age, on average and for different positions. Data correspond to the filtered database, i.e. without social sciences and particle physics, which explains the differences with Figures~\ref{fig:agedissem} and~\ref{fig:gradedissem}.}
\label{vulga_age_grade}
\end{center}
\end{figure*}

Results (1) and (3) represent a clear example of the usefulness of a regression study. Since age and position are strongly correlated, a simple study of the evolution of the proportion of active scientists with age is not concluding. Figure~\ref{vulga_age_grade} confirms that popularization activity decreases with age for all positions, but that scientists in higher hierarchical positions are much more active.

It should be noted that there are strong positive correlations between different types of dissemination. As argued previously, this shows that different activities do not compete for scientist time. Table~\ref{regression1} also show the specificities of the subdisciplines for the different dissemination activities. For example, mathematicians are much more active in teaching, less in industrial collaborations \ldots

Table~\ref{regression2} shows how scientists' characteristics influence the probability of being active in {\it all} the dissemination activities. In general, the effects are similar to those seen for single activities, but the effects are stronger. The main difference is the strong effect of academic activity, which is even stronger than for industrial collaborations alone. This further confirms that the different activities (academic and dissemination) are not competing but tend to reinforce mutually. Table~\ref{regression2} also shows the effects of age, gender, etc. on the probability of being inactive in all dissemination activities. These effects are consistent (i.e. opposite) with those seen for the probability of being active, except for the lack of significant effect of the academic activity.

\begin{table}[hp]
\centering
\small
\caption{Binomial regressions to explain dissemination activities for 3659 CNRS scientists (filtered database). ``Active'' means at least one action in the three-year period encompassed by our study (2004-2006). For popularization, this represents 1495 scientists (40.9\%), for industrial collaborations 2058 scientists (56.2\%) and for teaching 2373 scientists (64.9\%). These percentages are different from those of Table~\ref{propDS} because Social sciences are excluded from the filtered database. The explanatory variables are: sex, age, position, subdiscipline and Hirsch index ($h$) as bibliometric quantifier. The reference levels are: ``Condensed matter: structure'' for the subdiscipline and DR2 for the position. The columns give the coefficients of the fit, together with their significance (standard significance codes for the p-values have been used: ``***'' for $<0.001$,  ``**'' for $<0.01$, ``*'' for  $<0.05$, ``.'' for $<0.1$.)  The position ``DRCE'' is almost never significative because it includes less than 100 scientists (Table~\ref{tab:gradedissem}).}
\centerline{
  \begin{tabular}{|crp{4.89cm}|d{3}@{}l|d{3}@{}l|d{3}@{}l|d{3}@{}l|d{3}@{}l|}
    \hline
    &&Subdiscipline
    &\multicolumn{2}{c|}{press\rule[-0.9em]{0.cm}{2.7em}}
    &\multicolumn{2}{c|}{\parbox{1cm}{open days}}
    &\multicolumn{2}{c|}{
      \begin{tabular}[c]{@{}c@{}}active in\\ pop.\end{tabular}
}
    &\multicolumn{2}{c|}{
      \begin{tabular}[c]{@{}c@{}}active in\\ industrial\end{tabular}
}
    &\multicolumn{2}{c|}{\parbox{1.3cm}{\centering active in teaching}}
\\
    \hline
    &&(Intercept)&-1.8&*&-1.7&**&0.71&.&1.9&***&3.1&***\\
    \hline
    &&$h$      &0.037&**&-0.033&*&0.0020&&0.019&*&0.0067&\\
    &&age      &-0.057&***&-0.0010&&-0.027&***&-0.052&***&-0.078&***\\
    &&CR2      &-0.97&**&0.21&&-0.58&**&-1.5&***&-1.7&***\\
    &&CR1      &-0.28&&0.089&&-0.25&*&-0.72&***&-0.94&***\\
    &&DR2      &\multicolumn{10}{c|}{reference}\\
    &&DR1      &0.21&&-0.37&&0.24&.&0.70&***&0.31&*\\
    &&DRCE     &0.37&&-14&&0.26&&0.96&*&-0.15&\\
    &&active in popularization  &\multicolumn{2}{c|}{X}&\multicolumn{2}{c|}{X}&\multicolumn{2}{c|}{X}&0.40&***&0.39&***\\
    &&active in industrial collab.&0.37&*&0.25&.&0.47&***&\multicolumn{2}{c|}{X}&0.61&***\\
    &&active in teaching&0.44&**&0.42&**&0.60&***&0.40&***&\multicolumn{2}{c|}{X}\\
    &&male     &0.15&&-0.21&&-0.19&*&-0.08&&0.45&***\\
    \hline
    \multirow{5}{*}{{\begin{tabular}{@{}c@{}}Physical\\ sciences\end{tabular}}}
    &1&Mathematics                &0.31&&-1.6&**&-0.41&&-1.2&***&0.96&***\\
    &2&Physics, theory and method &1.1&.&-0.92&.&-0.37&&-0.64&*&0.06&\\
    &4&Atoms and molecules        &1.5&**&0.34&&0.097&&-0.18&&0.19&\\
    &5&Condensed matter: dynamics &0.5&&-0.017&&0.15&&0.21&&0.17&\\
    &6&Condensed matter: structure&\multicolumn{10}{c|}{reference}\\
    \hline
    \multirow{4}{*}{Engineering}
    &7&Information science          &1.3&*&-0.72&&-0.48&&1.5&***&2.0&***\\
    &8&Micro and nano-technologies  &1.2&*&-0.75&.&-0.11&.&2.5&***&0.69&*\\
    &9&Materials and structure      &1.4&*&-0.53&&-0.32&&1.9&***&2.4&***\\
    &10&Fluids and reactants        &1.4&*&-0.41&&0.032&&2.1&***&1.0&***\\
    \hline
    \multirow{6}{*}{Chemistry}
    &11&Super/macromolecular systems&-0.18&&-0.04&&-0.35&&0.92&***&0.80&**\\
    &12&Molecular architecture      &-1.4&&-1.1&*&-0.91&**&0.86&**&0.73&**\\
    &13&Physical chemistry          &-0.077&&-0.43&&-0.59&*&0.011&&0.46&.\\
    &14&Coordination chemistry      &-0.50&&-0.16&&-0.35&&1.0&***&0.027&\\
    &15&Materials chemistry         &0.83&&-0.8&.&-0.52&*&1.3&***&0.65&*\\
    &16&Biochemistry                &0.28&&-0.51&&-0.81&**&0.97&***&0.82&**\\
    \hline
    \multirow{4}{*}{\parbox{1.9cm}{\centering Earth\newline sciences,\newline astrophysics}}
    &17&Solar systems, universe          &2.3&***&0.56&&1.2&***&-1.1&***&-0.11&\\
    &18&Earth and earth plants           &1.6&**&-0.19&&0.63&*&-0.37&&1.9&***\\
    &19&Earth systems: superficial layers&2.3&***&-0.58&&0.76&*&0.17&&0.48&\\
    &20&Continental surface              &1.8&**&-0.82&&0.72&*&1.0&**&1.4&***\\
    \hline
    \multirow{10}{*}{Life sciences}
    &21&Molecular basis of life      &0.064&&-1.0&*&-1.1&***&1.1&***&1.1&***\\
    &22&Genomic organization         &-0.065&&-1.5&**&-1.0&***&0.12&&1.1&***\\
    &23&Cellular biology             &-1.6&&-0.97&*&-1.0&***&0.51&.&1.2&***\\
    &24&Cellular interaction         &0.31&&-1.4&**&-0.73&**&0.40&&1.4&***\\
    &25&Physiology                   &-0.17&&-1.9&**&-0.69&**&0.61&*&1.1&***\\
    &26&Development, evolution       &0.19&&-0.92&.&-0.70&**&0.37&&1.1&***\\
    &27&Behavior, cognition and brain&2.7&***&-1.6&**&0.56&*&0.78&**&1.8&***\\
    &28&Integrative vegetal biology  &0.53&&0.063&&-0.35&&0.57&*&0.79&**\\
    &29&Biodiversity, evolution      &2.1&***&-0.13&&0.4&&0.76&*&1.6&***\\
    &30&Therapy, pharmacology        &1.0&&-1.3&*&-0.58&*&1.1&***&0.85&**\\
\hline
\end{tabular}
}
\label{regression1}
\end{table}

\begin{table}[hp]
\centering
\caption{Binomial regressions to explain dissemination activities for 3659 CNRS scientists (filtered database). ``Active in all dissemination activities'' means at least one action in each of the activities over the three year period and ``Inactive'' no action in any of the three dissemination activities. The explanatory variables are: sex, age, position, subdiscipline and Hirsch index ($h$) as bibliometric quantifier. The reference levels are: ``Condensed matter: structure'' for the subdiscipline and DR2 for the position. The columns give the coefficients of the fit, together with their significance (standard significance codes for the p-values have been used: ``***'' 0.001 ``**'' 0.01 ``*'' 0.05 ``.'' 0.1.
)}
\begin{tabular}{|crp{9.4cm}|d{2}@{ }l|d{3}@{ }l|}
  \hline
  &&Subdiscipline
  &\multicolumn{2}{c|}{\parbox{1.5cm}{\centering\rule{0pt}{1eM} inactive\newline for all\rule[-1ex]{0pt}{1pt}}}
  &\multicolumn{2}{c|}{\parbox{1.5cm}{\centering active\newline for all}}\\
  \hline
  &&(Intercept)&-6.4&***& 1.5&**\\
  &&$h$&-0.01&&0.027&**\\
  &&age&0.10&***& -0.071&***\\
  &&CR2&2.3&***& -1.7&***\\
  &&CR1&1.1&***& -0.86&***\\
  &&DR2&\multicolumn{4}{c|}{reference}\\
  &&DR1&-0.71&**&0.60&***\\
  &&DRCE&-0.23&&0.20&\\
  &&male&-0.32&**& -0.15&\\
  \hline
  \multirow{5}{*}{{\begin{tabular}{@{}c@{}}Physical\\ sciences\end{tabular}}}
  &1&Mathematics&0.19&& -0.63&\\
  &2&Physics, theory and method&0.77&*& -0.32&\\
  &4&Atoms and molecules, lasers and optics&-0.02&& -0.25&\\
  &5&Condensed matter: organization and dynamics&0.19&&0.74&*\\
  &6&Condensed matter: structure&\multicolumn{4}{c|}{reference}\\
  \hline
  \multirow{4}{*}{Engineering}
  &7&Information science and technology&-1.9&**& 1.2&***\\
  &8&Micro and nano-technologies, electronics and photonics&-1.9&***& 1.4&***\\
  &9&Materials and structure engineering&-1.7&**& 1.7&***\\
  &10&Fluids and reactants: transport and transfer&-1.5&***& 1.7&***\\
  \hline
  \multirow{6}{*}{Chemistry}
  &11&Super and macromolecular systems, properties and functions&-0.79&*&0.55&\\
  &12&Molecular architecture synthesis&-0.25&&0.31&\\
  &13&Physical chemistry: molecules and environment&0.31&& -0.18&\\
  &14&Coordination chemistry: interfaces and procedures&-0.26&&0.31&\\
  &15&Materials chemistry: nanomaterials and procedures&-0.85&*&0.72&*\\
  &16&Biochemistry&-0.66&.&0.45&\\
  \hline
  \multirow{4}{*}{\parbox{1.9cm}{\centering Earth\newline sciences,\newline astrophysics}}
  &17&Solar systems and the universe&-0.06&& -0.40&\\
  &18&Earth and earth plants&-2.4&***&0.62&.\\
  &19&Earth systems: superficial layers&-1.9&**&0.34&\\
  &20&Continental surface and interfaces&-1.2&*& 1.9&***\\
  \hline
  \multirow{10}{*}{Life sciences}
  &21&Molecular basis and structure of life systems&-0.47&&0.41&\\
  &22&Genomic organization, expression and evolution&0.03&&0.087&\\
  &23&Cellular biology: organization and function&-0.39&&0.27&\\
  &24&Cellular interaction&-0.48&&0.26&\\
  &25&Molecular and integrative physiology&-0.74&*&0.34&\\
  &26&Development, evolution, reproduction and aging&-0.43&&0.14&\\
  &27&Behavior, cognition and brain&-1.7&***& 1.7&***\\
  &28&Integrative vegetal biology&-0.32&&0.61&.\\
  &29&Biodiversity, evolution and biological adaptation&-1.4&**& 1.6&***\\
  &30&Therapy, pharmacology and bioengineering&-0.46&& 0.80&*\\
  \hline
\end{tabular}
\label{regression2}
\end{table}

To interpret the correlations between academic and dissemination activities, one has to take into account that our regression models quantify academic achievements in two different ways: the position and the bibliometric indicator $h$. Since we include both in our regressions, the effect of $h$ is considered ``all other things being equal'', i.e. only within each position. Therefore, the lack of influence of $h$ for the probability of being active in popularization (third column of Table~\ref{regression1}) is not in contradiction with our previous findings (Table~\ref{hyDS} for example). Indeed, since positions are strongly correlated with $h$ and higher positions are much more active in popularization (Table~\ref{regression1}), our regression shows that that popularization activity is more influenced by the hierarchical position than by the bibliometric indicators within each category\footnote{If positions are omitted from the regressions, h becomes strongly significative.}. The opposite is observed for the probability of participating in ``open days'' events (second column of Table~\ref{regression1}). Here, position is not relevant (none affects significantly the participation even if a definite trend exists) but Hirsch index is. A similar (but in the opposite direction) conclusion can be drawn for popularization in the press (first column of Table~\ref{regression1}): position is only slightly relevant (except for the youngest scientists, which rarely participate), but $h$ is very important, as is scientist's age. Briefly, the main influences for the other types of actions are the following: contracts with industrial partners are strongly (positively) influenced by the  position, Hirsch index and (decreasing) age. Instead, popularization through television and conferences or industrial collaborations through patents or licenses only depend on position and age and not bibilometric indicators. Finally, participation in popularization in schools decreases with age but is unaffected by the position or academic activity.

\section{Interpreting our results}

What can we learn from this statistical study of dissemination activities? We will examine different hypothesis and compare them to our findings. Our interpretations are centered on popularization practices. We will deal with industrial collaborations in future work.

\begin{itemize}

\item\textbf{H0: Dissemination is done by ``those who are not good
enough for an academic career''.} If we admit that bibliometric
indicators are a good proxy for ``being good enough for an academic
career'', then our study clearly invalidates this hypothesis. First,
randomly chosen active scientists have higher academic indicators than
inactive ones. Second, all other things being equal, the probability
of disseminating increases with academic position
(Table~\ref{regression1}). Furthermore, better academic records
increase the probability of being active in industrial collaborations.

\item\textbf{H1: Dissemination is done by people close to retirement.}
Our study has also shown that this hypothesis is
incorrect. Table~\ref{tab:agedissem} shows that scientists close to
retirement are less active than average. Our statistical also shows
that, as their age increases, scientists become less active in
dissemination, all other things being equal.

\item\textbf{H2: Popularization is driven by demand upon the
scientific elite.} This hypothesis assumes that popularization is
driven by an external demand (institutions or journalists). Then, the
scientific elite, with higher bibliometric indices, is more visible
from outside the scientific community, and is therefore more
solicited. Our data supports this interpretation: scientists engaged
in the type of popularization actions mostly driven by demand (radio,
television, press, and to a lesser extent, conferences) have a higher
$h_y$ than average (Table~\ref{demand}). Instead, scientists
performing the popularization activities that are mostly driven by the
offer, and symbolically less important (open days, school conferences,
web sites...) have a slightly lower average $h_y$ than the other
scientists (Tables~\ref{regression1} and~\ref{demand}).

\item\textbf{H3: Active personality.} We have repeatedly pointed out
that being active in one activity is positively correlated to being
active in the others. This suggests that the correlations observed
could arise from some internal characteristic of the scientists
involved, call it ``active personality'' or ``intellectual
capacity''. For example, it could be argued that popularization is
intellectually demanding, for it is difficult to explain complex
scientific issues in simple terms. Therefore, good popularization
demands a deep understanding of the subject, as anyone preparing
lectures has experienced. This ``intellectual capacity'' would in turn
generate higher academic records. Alternatively, one could interpret
an ``active personality'' as one able to ``sell'' his work, both to
journalists and to those colleagues in charge of refereeing papers and
citing them. The ``active personality'' argument is supported by the
the fact that scientists active in all dissemination activities are
also very active academically speaking.

\item\textbf{H4: Social and cognitive hierarchies.} The observed
correlations between position and dissemination activities can also be
understood by referring to sociological studies of scientific
communities. Terry Shinn~\cite{shinn} studied a French physics lab for
several years, looking for correlations between hierarchical positions
and cognitive work. He noticed a clear work division between junior
and senior scientists. Junior staff devote most of their time to
experiments or "local" questions. By ``local'', Shinn means questions
focused on particular points : a single experiment, a thorough
investigation of a very precise subtopic \ldots. In contrast, senior
scientists devote most of their time to "general" questions, i.e. how
the local results can be inserted into global theoretical or
conceptual frameworks. They also spend much time establishing and
maintaining social networks both inside and outside the scientific
community. Both these activities are clearly more in line with
dissemination activities, which demand putting scientific problems
into perspective. Moreover, senior scientists generally have a team of
junior researchers working with them, which is active even when
seniors are outside of the lab disseminating... One could also argue
that the scientific elite is able to transform its symbolic
capital~\cite{bourdieu}, gained in the academic arena, into public
arenas, thus popularizing not only on issues directly related to their
own domain but on virtually any issue. This would lead to a
correlation between higher popularization activity and academic
records, as for H2. In an old study of popularization practices of
CNRS scientists, Luc Boltanski~\cite{boltanski} has observed that
Senior scientists ("Directeur de Recherche") have the legitimacy to
speak to the public in the name of the institution. Instead,
scientists in the lowest positions can only express their own point of
view, and popularization is mostly seen as a waste of time or a
personal occupation.

\item\textbf{H5: Benefits of dissemination}. It is the reverse
causality from H2 and H4. Dissemination activities compel scientists
to open up their horizon, to discuss with people having other points
of view on their research topics, giving new insights, contacts, which
could improve their academic research. Actually, H2, H4 and H5 could
act together in a reinforcing way. It seems difficult to argue that
this effect is dominating, but it could contribute to the observed
correlations, mainly in the case of industrial collaborations which
strongly correlate with higher academic indicators.
\end{itemize}

As a summary, it is likely that the strong correlations observed between dissemination and academic activity result from the cumulative effects of H2-H5.

\begin{table}[hp]
\begin{center}
\caption{Differences in scientific activity --- as measured by the normalized Hirsch index --- for different subpopulations, characterized by the type of their popularization activities. The first group refers to prestigious activities regulated mainly by the outside demand, whereas the second group gathers less prestigious activities mainly driven by scientists' offer. The p-values give the statistical significativity of the differences. They are obtained by a standard 'Welch Two Sample t-test'. Note that the differences for the first group are highly significant even if the number of scientists active in those activities is quite low: between 220 (radio/television) and 430 (public conferences).} 
\begin{tabular}{|l|c|c|c|c|}
\hline
type of action & \multicolumn{2}{c|}{$h_y$}& p-value\\
 & active & inactive & \\
\hline
Press         &                0.82 & 0.70 & 1.8 $10^{-6}$ *** \\                
Radio, Television        &                0.81 & 0.70 & 5.5 $10^{-5}$ *** \\
Public Conference        &                0.75 & 0.70 & 0.019 * \\
\hline
School conference        &                0.73 & 0.71 & 0.18 \\
Open days        &                0.68 & 0.71 & 0.11 \\
Web sites        &                0.73 & 0.71 & 0.55 \\
\hline
\end{tabular}
\label{demand}
\end{center}
\end{table}

\section{Are dissemination activities good for the career?}

It is commonly recognized that scientists engaged in dissemination do
not get much reward, and that their involvement can even be bad for
their career~\cite{royal2006}. In France, the CNRS director stated
recently the importance of taking into consideration
``scientific culture popularization actions'' for the
evaluation of researchers: ``one must insist that they give
equal importance to scientific work and to activities related to the
popularization and dissemination of scientific culture: participations
in ``open doors'' events, or the publication in
magazines or other popularization works, in events organized for
non-specialized audiences, newspaper articles or TV appearances,
etc.'' (letter sent to CNRS scientists in 2005, our
translation). In the document that was supposed to steer his long-term
policy, the ``Multi-year action plan''~\cite{CAPCNRS},
the CNRS thus declares that: ``If current [evaluation] practice
is suitable for the purpose of evaluating academic research, the same
cannot be said for interdisciplinary activities and for other facets
of scientific work: transfer of scientific knowledge, teaching and
popularization. Consequently, the work by CNRS researchers who choose
to engage in these activities, which are very necessary for the CNRS,
is not adequately acknowledged and researchers are therefore reluctant
to proceed in this direction.''

Thanks to our large database, we are able to study statistically the
influence of dissemination activities on the promotions of CNRS
researchers to senior positions (``Directeur de Recherche'') over the
2004--2006 period. Table~\ref{promo} shows the results of our
regression analysis, for all CNRS disciplines and for each discipline
separately. It turns out that dissemination activities are {\it not}
bad for scientists careers. They are not very good either: the effects
are generally weak, but positive, and rarely significative. The
detailed study by discipline shows that the overall positive effect of
popularization arises mainly from its recognition in Life sciences and
the positive effect of teaching from Chemistry. However, it is
interesting to note that all dissemination activities influence
positively promotions for most of other disciplines, even if their
effects are not significative.

Overall, two characteristics have strong effects: academic activity (h or the number of papers) and age (the ``optimal'' age for becoming DR2 is 46.6 years, for DR1 52.4). For DR1 promotion, there is a small (and positive) effect from industrial collaborations.

\begin{table}[hp]
\centering
\small
\caption{Binomial regressions to explain promotions to senior positions (from CR1 to DR2) on the 586 candidates and 179 promotions from all scientific disciplines of our filtered database. The last column shows the regressions for the DR2 to DR1 promotion, on 376 candidates and 67 promotions from all scientific disciplines of our filtered database. The explanatory variables are: sex, age, subdiscipline (not shown to simplify since none is significative) and $h$ as bibliometric quantifier (except for Engineering, where the number of articles accounts much better for the promotions,  see~\cite{testh}). The columns give the coefficients of the fit for each scientific domain, together with their significance. Standard significance codes for the p-values have been used:  0 ``***'' 0.001 ``**'' 0.01 ``*'' 0.05 ``.'' 0.1.}
\centerline{
\begin{tabular}{|l||d{3}@{ }l|d{3}@{ }ld{3}@{ }ld{3}@{ }ld{4}@{ }ld{2}@{ }l|d{3}@{ }l|}
\hline
&\multicolumn1c{all}&&\multicolumn1c{Physics}&&\multicolumn1c{ENG}&&\multicolumn1c{Earth}&&\multicolumn1c{Chemistry}&&\multicolumn1c{Life}
&&\multicolumn1{c}{DR1}&\\
\hline
(Intercept)       &-41.2 &   &-78.2 &   &-71.8  &*&-35.41&.&-42.94 & &3.53   &  &-49.6&**\\
\hline
act pop                                                &0.4   &.  &0.01  &   &0.068  & &-0.82 & &0.38   & &0.91   &* &-.29 &\\
act indus                            &0.12  &         &0.88  &   &-1.22  & &-0.085& &0.06   & &0.14   &  &.65  &.\\
act teach                                        &0.79  &** &0.76  &   &1.66   & &1.08  & &1.06   &*&0.41   &  &.33  &\\
\hline
male              &-0.23 &   &-0.99 &   &0.72   & &0.5   & &-0.68  & &0.23   &  &-.36 &\\
$h$ (or art for ENG)&0.13  &***&0.26  &***&0.038  &*&0.12  &.&0.11   &*&0.15   &**&.106 &***\\
age             &1.12  &***&2.48  &*  &3.2    &*&1.26  & &0.96   &.&0.46   &  &1.78 &**\\
age$\null^2$    &-0.012&***&-0.026&*  &-0.036 &*&-0.012& &-0.0096&.&-0.0057&  &-0.017&**\\
\hline
\end{tabular}
}

\label{promo}
\end{table}

\section{Discussion, Conclusions}

Our statistical study on the correlations between dissemination
activities and academic records of more than 3500 scientists from most
disciplines has allowed us to establish several facts. First, we have
clearly shown that scientists engaged in dissemination are more active
academically, formally refuting the common idea that ``dissemination
activities are carried out by those who are not good enough for an
academic career''~\cite{royal2006}. We have even shown that some
prestigious activities (presse, radio and television) are mostly
carried out by the scientific ``elite'' in academic terms. One can
certainly criticize the idea that bibliometric indicators do account
properly for the academic quality of
scientists~\cite{liu,brooks,kostoff,leydes,testh}. However, those who
argue that dissemination activities are carried out by the ``worst''
scientists usually do accept this definition of scientific
quality. Therefore, our paper should convince them that they are
wrong.

This was the ``easy'' part of the discussion. The interpretation of
our results is otherwise not easy, as there have been few qualitative
studies on the perception by scientists of popularization or teaching
practices. Concerning relations with industries, a group of the
Catholic University of Leuven (K.U. Leuven, Belgium) examined the
academic records of 32 scientists ``inventors'' in their university~\cite{louvain}. Their data suggest a ``reinforcing or positive
spillover effect on scientific performance from engaging in technology
development efforts'', consistent with our findings in a much larger
sample. A recent study has investigated the factors that predict
scientists' intentions to participate in public engagement~\cite{poliakoff}. Thanks to a questionnaire distributed to academic
staff and postgraduates, it was found that the main reasons why
scientists decided not to participate in public engagement activities
are the following : they had not participated in the past (a result
consistent with our former finding in~\cite{jcom}), they have a
negative attitude toward participation (it is seen as 'pointless' or
'unenjoyable'), they feel they lack the skills and finally that they
do not believe that their colleagues participate in such activities,
which is interpreted as a signal of relative irrelevance of this
activity. Notably, the lack of time or career recognition are not seen
as important determinants of participation. Clearly, more qualitative
studies on the relations of the scientific milieu and dissemination
practices are needed. Here, we limit our discussion to a few issues:

\begin{itemize}

\item\textbf{What have we learnt about relations of scientists and
popularization?}  Our study suggests that popularization is mostly an
activity of the (academically speaking) scientific ``elite''. Our
finding agrees with data from the Royal Society survey~\cite{royal2006}, which shows that higher positions popularize more,
the differences being even larger than in our database : senior staff
is active at 86\%, while junior staff is active at a mere 14\%. The
following comment follows : "The seniority finding is borne out by the
qualitative research which found that young researchers keen to climb
the research career ladder were focused on research and publishing
and/or felt that they needed more experience before they could engage
with those outside their research community." We may add (H4 above)
that senior activities are more in line with dissemination that those
carried out by junior staff. The question that our findings rise is
then: why does a significant fraction of the scientific community feel
that ``only bad scientists'' popularize? Is it a problem of jealousy
for colleagues that manage to present their results to a wide
audience? Is that because, cognitively speaking, creating knowledge is
judged more important than disseminating it, as suggested by Shinn~\cite{shinn}? This would imply that scientists are still prisoners of
the ``diffusion model''~\cite{weigold}, which ignores that to
disseminate knowledge, one has to recreate it altogether, a creative
and difficult endeavour.

Second, qualitative interviews indicate that many reasons push
scientists to engage in popularization. In private discussions,
popularizers acknowledge that one of the main reasons is the pleasure
of interacting with the public, of going out of the
lab~\cite{argentina}. For the Royal Society study~\cite{royal2006},
i.e. in a more official environnment, the strongest reason given to
justify popularization is ``informing the public''. We can wonder
whether scientists are still prisoners of the so-called deficit
model~\cite{weigold}. This is an old model for scholars of the Science
Studies field, dating back to 1960. It insists on the teaching of
elementary scientific facts and methods to the public. Listening to
the public seems important to only a few percent of the scientists
interviewed in the UK~\cite{royal2006}. However, this idea should be
one of the strongest with a more ``generous'' vision of
the public in mind~\cite{jmll_pus,wagner}. Scientists also seem to
ignore the numerous criticisms to the deficit model: the relation
between the knowledge of scientific facts and its appreciation is
empirically unsolved, the knowledge of the ``facts'' of
science taken out of their context is more alienating than it is
informative... It is also important to establish links in the other
way, where scientists learn from
society~\cite{irwin,bj_pus,bbv_pus}. The culture of the scientific
milieu seems far from these ideas at the moment.

\item\textbf{What do we learn about relations between science and
society?}  Our study shows that, even in the institution hosting the
most fundamental sciences, roughly half of the scientists are in close
contact with society, i.e. popularize or look for funding outside the
academic sphere. This could worry some ``fundamentalists'' which think
that science should be isolated from society needs, because society
can only perturb science. For example, the recent French movement
``Sauvons la recherche'' wrote in its final document that ``science
can only work by developing its own questions, protected from the
emergency and the deformation congenital to social and economics
worlds''. However, our result will not surprise scholars in Science
Studies, as they know that the ``ivory tower'' of science never
existed. Scientists have always been connected with society, from
which they depend for funding (see, among many others,~\cite{biagioli,pestre,action}. Even the most fundamental physical theories such as relativity~\cite{galison} and quantum
mechanics~\cite{maze} have been inspired by applications.

\item\textbf{Dissemination and career} Another contradiction between
our study and common views among scientists is the idea that
``dissemination activities are negative for the career''. For example,
20\% of scientists of the Royal Society survey~\cite{royal2006}
answered that scientists who engage in popularization are viewed less
well by their peers. Here, we have shown that promotion is mainly
determined by academic indicators, dissemination activities being
marginal, counting only for specific disciplines (chemistry for
teaching, life sciences for popularization). There is however no
negative effect. How can we explain the common opposite idea that
pervades the scientific community?

\end{itemize}

We have started our paper by the proclamation of many prestigious institutions that dissemination activities are priorities. We have shown that these activities are carried out by academically active scientists, that receive no reward for their engagement. We feel that institutions' duty is now to invent ways of evaluating and rewarding the active scientists.

\bibliography{biblio}

\begin{thebibliography}{34}
\providecommand{\natexlab}[1]{#1}
\providecommand{\url}[1]{\texttt{#1}}
  \providecommand{\doi}[1]{\url{http://dx.doi.org/#1}}

\bibitem[Allix (2007)]{SSCNRS}
J.P. Allix.
\newblock \emph{Sciences et société en mutations}.
\newblock Editions du CNRS, 2007.

\bibitem[B.~Van~Looy (2006)]{louvain}
K.~Debackere B.~Van~Looy, J.~Callaert.
\newblock Publication and patent behaviour of academic researchers :
  Conflicting, reinforcing or merely co-existing ?
\newblock \emph{Research Policy}, 35:\penalty0 596, 2006.

\bibitem[Bensaude-Vincent (2001)]{bbv_pus}
Bernadette Bensaude-Vincent.
\newblock A genealogy of the increasing gap between science and the public.
\newblock \emph{Public Understanding of Science}, 10\penalty0 (1):\penalty0
  99--113, 2001.
\newblock \doi{10.1088/0963-6625/10/1/307}.

\bibitem[Biagioli (1993)]{biagioli}
M.~Biagioli.
\newblock \emph{Galileo, Courtier: The Practice of Science in the Culture of
  Absolutism}.
\newblock University of Chicago Press, 1993.

\bibitem[Boltanski and Maldidier (1969)]{boltanski}
L~Boltanski and P~Maldidier.
\newblock La vulgarisation scientifique et ses agents, 1969.

\bibitem[Bourdieu (1984)]{bourdieu}
P.~Bourdieu.
\newblock \emph{Distinction: A Social Critique of the Judgement of Taste}.
\newblock London, Routledge, 1984.

\bibitem[Brooks (1996)]{brooks}
T.A. Brooks.
\newblock Evidence of complex citer motivations.
\newblock \emph{Journal of the American Society for Information Science},
  37:\penalty0 34, 1996.
\newblock \doi{10.1002/(SICI)1097-4571(198601)37:1<34::AID-ASI5>3.0.CO;2-0}.

\bibitem[Cheveigné (2007)]{cheveigne}
S.~Cheveigné.
\newblock Private communication, 2007.

\bibitem[CNRS (2004)]{CAPCNRS}
CNRS.
\newblock Dissemination is one of the official duties of cnrs researchers.
  furthermore, in its long-term policy engagement with the french state, cnrs
  recognizes the importance of science dissemination and its proper evaluation,
  2004.
\newblock \url{http://www2.cnrs.fr/sites/band/fichier/3f1d5636c99a3.htm}.

\bibitem[Galison (2004)]{galison}
P.~Galison.
\newblock \emph{Einstein's Clocks, Poincare's Maps}.
\newblock Sceptre, 2004.

\bibitem[Hartz and Chappell (1997)]{sagan}
J.~Hartz and R.~Chappell.
\newblock \emph{Worlds Apart: How the Distance Between Science and Journalism
  Threatens America's Future}.
\newblock Freedom Forum First Amendment Center, Nashville, T N, 1997.

\bibitem[Hilgartner (1990)]{hilgartner}
S.~Hilgartner.
\newblock The dominant view of popularization: Conceptual problems, political
  uses.
\newblock \emph{Social Studies of Science}, 20:\penalty0 519, 1990.
\newblock \doi{10.1177/030631290020003006}.

\bibitem[Hirsch (2005)]{hirsch}
J.E. Hirsch.
\newblock An index to quantify an individual's scientific research output.
\newblock \emph{Proc. Natl. Acad. Sciences}, 102:\penalty0 16\,569--16\,572,
  2005.
\newblock \url{http://www.pnas.org/cgi/content/abstract/102/46/16569}.

\bibitem[Hoddeson et~al. (1992)]{maze}
Lillian Hoddeson, Ernst Braun, and Spencer~Weart (Eds).
\newblock \emph{Out of the Crystal Maze: Chapters from the History of Solid
  State Physics}.
\newblock Oxford University Press, 1992.

\bibitem[Iglesias and Pecharromán (2007)]{iglesias}
J.E. Iglesias and C.~Pecharromán.
\newblock Scaling the h-index for different scientific {ISI} fields.
\newblock \emph{Scientometrics}, 2007.
\newblock \url{http://arxiv.org/abs/physics/0607224}.

\bibitem[Irwin and Wynne (1996)]{irwin}
Alan Irwin and Brian Wynne.
\newblock \emph{Misunderstanding Science? The Public Reconstruction of Science
  and Technology}.
\newblock Cambridge University Press, 1996.

\bibitem[Jensen (2005)]{vulganature}
Pablo Jensen.
\newblock Who's helping to bring science to the people?
\newblock \emph{Nature}, 434:\penalty0 956, 2005.
\newblock \doi{10.1038/434956a}.

\bibitem[Jensen and Croissant (2007)]{jcom}
Pablo Jensen and Yves Croissant.
\newblock {CNRS} researchers' popularization activities: a progress report.
\newblock \emph{Journal of Science Communication}, October 2007.
\newblock
  \url{http://jcom.sissa.it/archive/06/03/Jcom0603(2007)A01/Jcom0603(2007)A01_%
fr.pdf}.

\bibitem[Jensen, Rouquier, and Croissant (2009)]{testh}
Pablo Jensen, Jean-Baptiste Rouquier, and Yves Croissant.
\newblock Testing bibliometric indicators by their prediction of scientists
  promotions.
\newblock \emph{Scientometrics}, 78\penalty0 (3), March 2009.
\newblock ISSN 0138-9130.
\newblock \doi{10.1007/s11192-007-2014-3}.

\bibitem[Jurdant (1993)]{bj_pus}
B.~Jurdant.
\newblock Popularization of science as the autobiography of science.
\newblock \emph{Public Understanding of Science}, 2:\penalty0 365, 1993.
\newblock \doi{10.1088/0963-6625/2/4/006}.

\bibitem[Kostoff (1998)]{kostoff}
R.~N. Kostoff.
\newblock The use and misuse of citation analysis in research evaluation.
\newblock \emph{Scientometrics}, 43\penalty0 (1):\penalty0 27, September 1998.
\newblock \doi{10.1007/BF02458392}.
\newblock \url{http://www.akademiai.com/content/e445v73tgth7r702}.

\bibitem[Latour (1988)]{action}
B.~Latour.
\newblock \emph{Science in Action: How to Follow Scientists and Engineers
  through Society}.
\newblock Harvard University Press, 1988.

\bibitem[Leydesdorff (1998)]{leydes}
L.~Leydesdorff.
\newblock Theories of citation?
\newblock \emph{Scientometrics}, 43:\penalty0 5, 1998.
\newblock \doi{10.1007/BF02458391}.

\bibitem[Liu (1993)]{liu}
Mengxiong Liu.
\newblock The complexities of citation practice: a review of citation studies.
\newblock \emph{Journal of Documentation}, 49\penalty0 (4):\penalty0 370--408,
  1993.
\newblock ISSN 0022-0418.

\bibitem[Lévy-Leblond (1992)]{jmll_pus}
Jean-Marc Lévy-Leblond.
\newblock About misunderstandings about misunderstandings.
\newblock \emph{Public Understanding of Science}, 1:\penalty0 17, 1992.

\bibitem[Pestre (2003)]{pestre}
D.~Pestre.
\newblock \emph{Science, argent et politique}.
\newblock INRA Editions, 2003.

\bibitem[Poliakoff and Webb (2007)]{poliakoff}
Ellen Poliakoff and Thomas~L. Webb.
\newblock What factors predict scientists' intentions to participate in public
  engagement of science activities?
\newblock \emph{Science communication}, 29:\penalty0 242, 2007.

\bibitem[Pérez, Kreimer, and Jensen (2008)]{argentina}
A.~Pérez, P.~Kreimer, and Pablo Jensen.
\newblock Scientists motivations for popularizing science in argentina.
\newblock \emph{In preparation}, 2008.

\bibitem[Shermer (2002)]{shermer}
M.B. Shermer.
\newblock Who's helping to bring science to the people?
\newblock \emph{Social Studies of Science}, 32:\penalty0 489, 2002.

\bibitem[Shinn (1988)]{shinn}
Terry Shinn.
\newblock Hi\'erarchies des chercheurs et formes des recherches.
\newblock \emph{Actes de la recherche en sciences sociales}, 74:\penalty0 2,
  1988.

\bibitem[Society (2006)]{royal2006}
Royal Society.
\newblock Factors affecting science communication: a survey of scientists and
  engineers, 2006.
\newblock \url{http://www.royalsoc.ac.uk/page.asp?id=3180}.

\bibitem[Wagner (2007)]{wagner}
Wolfgang Wagner.
\newblock Vernacular knowledge.
\newblock \emph{Public Understanding of Science}, 16\penalty0 (1):\penalty0
  7--22, 2007.
\newblock \doi{10.1177/0963662506071785}.

\bibitem[Web of Science]{wos}
Web of Science.
\newblock \url{http://scientific.thomson.com/index.html}.

\bibitem[Weigold (2001)]{weigold}
Michael~F. Weigold.
\newblock Communicating science.
\newblock \emph{Science Communication}, 23\penalty0 (2):\penalty0 164--193,
  2001.
\newblock \doi{10.1177/1075547001023002005}.

\end{thebibliography}
\bibliographystyle{plainnat2}

\newpage
\section{Appendix 1 : A summary of popularization activities of CNRS scientists}

To draw up their annual report, researchers must specify the type of popularization activities that they have performed. The following table displays a distribution of types of activities according to the different scientific departments of the CNRS for the 2006 data. The categories are chosen by the scientists themselves. Most categories names in the Table are self-explanatory. 'Associations' refer to popularization actions taken to help associations understand scientific aspects of their activity (think of patients or astronomical associations). 'Schools' refers to actions taking place in schools. 'Web' to popularization sites on the Web.

It is interesting to analyze the misrepresentation of certain disciplines for each type of activity. For instance the over-representation of Social Sciences researchers in Radio/Television and, to a lesser extent, in activities involving associations, the press and conferences. Not surprisingly, these researchers are by far under-represented in ``open door'' events. On the other hand, their weak presence in schools is food for thought for the community. The Nuclear physics, Chemistry and Engineering departments are over-represented in ``open door'' activities, which are relatively scarce in Life Sciences. These departments are rather absent from actions involving the press, radio or publishing.

\begin{tabular}{lcccccccc}
\hline
 & Chemistry & Nuclear & Earth & Life & Social & Engineering & Physics & Info tech \\
\hline
Other & 0.14 & 0.11 & 0.06 & 0.10 & 0.05 & 0.10 & 0.10 & 0.10 \\
Conference & 0.18 & 0.24 & 0.28 & 0.19 & 0.30 & 0.20 & 0.24 & 0.23 \\
Exhibition & 0.09 & 0.08 & 0.06 & 0.05 & 0.06 & 0.11 & 0.11 & 0.09 \\
Associations & 0.02 & 0.03 & 0.05 & 0.05 & 0.05 & 0.04 & 0.04 & 0.02 \\
Schools & 0.14 & 0.14 & 0.09 & 0.14 & 0.02 & 0.13 & 0.11 & 0.10 \\
Books/CD Rom & 0.03 & 0.02 & 0.04 & 0.03 & 0.04 & 0.03 & 0.04 & 0.05 \\
Open doors & 0.16 & 0.09 & 0.08 & 0.09 & 0.01 & 0.13 & 0.12 & 0.13 \\
Press & 0.13 & 0.14 & 0.15 & 0.18 & 0.19 & 0.13 & 0.13 & 0.14 \\
Radio/Television & 0.06 & 0.06 & 0.14 & 0.14 & 0.22 & 0.08 & 0.07 & 0.06 \\
Web & 0.06 & 0.08 & 0.05 & 0.03 & 0.05 & 0.05 & 0.05 & 0.08 \\
\hline
\end{tabular}

\newpage
\section{Appendix 2 : Obtaining reliable bibliometric indicators for several thousand scientists}

In the following, we detail our procedure to obtain a large but reliable sample ($\simeq~3\,500$ records) of bibliometric indicators (number of publications, citations and $h$ index). The difficulty lies in the proper identification of the publications of each scientist. Two opposite dangers arise. The first one consists in including extra publications because the request is not precise enough. For example, if only surname and name initials are indicated to WoS, the obtained list may contain papers from homonyms. The second one consists in missing some papers. This can happen if scientists change initials from time to time, or if the surname corresponds to a woman who changed name after marriage. But this can also happen when one tries to be more precise to correct for the first danger, by adding other characteristics such as scientific discipline or French institutions for CNRS scientists. The problem is that both the records and the ISI classifications are far from ideal: the scientific field can be confusing for interdisciplinary research, the limitation to French institutions
incorrect for people starting their career in foreign labs, etc.

Basically, our strategy consists in guessing if there are homonyms (see below how we manage to get a good idea on this). If we think there are no homonyms, then we count all papers, for any supplementary information (and the resulting selection) can lead to miss some records. If we guess that there are homonyms, then we carefully select papers by scientific domain and belonging to French institutions. After all the bibliometric records have been obtained in this way, we filter our results to eliminate ``suspect'' records by two criteria: average number of publications per year and scientist's age at the first publication.

\subsection{Evaluate the possibility that there exist homonyms}
\label{homonyms-probability}
For this, compute the ratio of the number of papers found for the exact spelling
(for example JENSEN P.) and all the variants proposed by WoS (JENSEN P.*,
meaning P.A. P.B., etc.). If this ratio is large (in our study, larger than .8),
then the studied surname is probably not very common and the author might be the
single scientist publishing. To get a more robust guess, we use the scientist's
age. We look at the total number of papers and compare it to a ``maximum''
normal rate of publishing, taken to be 6 papers a year. If the publishing rate is smaller than our threshold,
this is a further indication that there is a single scientist behind all the
records. Actually, our strategy can be misleading only when there are only
homonyms with the same initial {\it and} all the homonyms have published very
few papers.

\subsection{Obtain the bibliometric records}
\label{obtain-biblio-record}
\paragraph{No homonyms}
If we guess that there is a single scientist behind the publications
obtained for the surname and initials (which happens for about 75\% of
the names), we record the citation analysis corresponding to all
associated papers.

\paragraph{Homonyms}
If we estimate that there are homonyms, we try to eliminate them by using
supplementary data we have. We refine the search by scientific field (``Subject
category'' in WoS terms, but one can select only one) and by selecting only
French institutions\footnote{Unfortunately, WoS allows the selection to be made
  only on the institutions of all coauthors as a whole. So we might retain
  articles of homonyms that have coauthored a paper with a French scientist.}.

\subsection{Eliminate suspect records}
Finally, once all the data has been gathered according to the preceding
steps, we eliminate ``suspect'' results by two criteria related to the scientist's age. For a record
to be accepted, the age of the first publication has to be between 21 and 30 years, and the average number of publications per year between 0.4 and 6. After this filtering process, we end up with 3659 records out of the 6900 initial scientists, i.e. an acceptance rate of 53\%.

Can we understand why half of the records are lost? First of all, let us detail
how the different filters eliminate records. Deleting scientists who published
their first paper after 30 years old eliminates 1347 ``suspect'' names, which
are probably related to errors or missing papers in the WoS database, to married
women for whom me miss the first papers published under their own surname and to
people who started their career in non French institutions and had homonyms.
Deleting scientists who published their first paper before 21 years old
eliminates 1235 additional ``suspect'' names, which are probably related to
errors in the WoS database, to scientists with older homonyms which we could not
discriminate. Deleting scientists whose record contained less than .4 papers per
year in average leads to the elimination of 121 names. These wrong records can
be explained by the method missing some publications, as in the case ``first
publication after 30 years old''.  Deleting scientists whose record contained
more than 6 papers per year in average leads to the elimination of further 178
names. These wrong records can be explained by the presence of homonyms we could
not discriminate. Finally, to make our database more robust, we decided to
eliminate records suspect of containing homonyms even after selection of
discipline and institution. This is done by eliminating the 359 scientists for
which the number of papers kept after selection is smaller than 20\% of the
total number of papers for the same surname and initials. In those cases, we do
not trust enough our selection criteria to keep such a fragile record.

\subsection{A robust bibliometric database}
In summary, our method leads to a reliable database of around 3500 scientists from all ``hard'' scientific fields. It only discriminates married women having changed surname. It also suffers from the unavoidable wrong WoS records\footnote{For a noticeable fraction of scientists, WoS records start only in the 1990s, even if there exist much older publications, which can be found for example by Google Scholar.}. We stress that the main drawback of the elimination of half the records is the resulting difficulty in obtaining good statistics. But at least we are pretty sure of the robustness of the filtered database. 

Our filtering criteria are based on homonym detection, age of first publication and publication rate. The first criteria correlates only with scientist's surnames, therefore we can expect that it introduces no bias except for married women. Actually, there is a lower woman proportion after filtering: 24.9\% women in the 3659 selection, against 29.6\% in the 6900 database. This is consistent with the preferential elimination of married women who changed surnames and have an incomplete bibliographical record. 

The two other criteria could discriminate scientific disciplines with lower publication rates or underrepresented in WoS. For example, we see in the following table that more scientists from the Engineering Department have been eliminated in the filtering. The mean age is somewhat lower in the filtered database (46.4 years) to be compared to 46.8 in the whole dataset, probably because the records from older scientists have a higher probability of containing errors.

However, overall, the filtered database is very similar to the initial one. For example, the percentage of candidates to senior positions is 16.0\% in the 3659 selection, against 16.4\% in the 6900 database, and the respective promoted percentages are 4.9\% and 5.0\%. The proportions of scientists from each position is also similar : none of the small differences between the filtered and unfiltered values is statistically significant.

As noted previously, the robustness of our filtered database is validated by the significantly better indicators found for scientists in higher positions. An even stronger test (because the effect is subtler) resides in testing the correlations of the scientist's age at his(her) first publication with several variables : age, position, subdiscipline and gender. We find
 a progressive decrease of the age of first publication when a scientist has a higher position (all things being equal, for example scientist's age), an effect that is intuitively appealing but certainly small. The fact that we can recover such a subtle effect is a good indication of the robustness of our procedure. We also recover the intuitive effect of scientist's age (older scientists have begun their career later). The gender effect (men publish their first paper 2 months later than women, all other things being equal) is more difficult to interpret, since it mixes many effects : our discrimination (in the filtering procedure) of married women, the unknown effects of marriage and children on scientists' careers, etc.

\end{document}